\documentclass[journal=jacsat,manuscript=article]{achemso}

\usepackage{chemformula} %
\usepackage[T1]{fontenc} %

\usepackage{bm}
\usepackage{mhchem}
\usepackage{physics}
\usepackage{cleveref}
\usepackage{cuted}
\usepackage{dirtytalk}
\usepackage{mathtools}
\usepackage{lineno}

\renewcommand{\hat}{\widehat}
\renewcommand{\tilde}{\widetilde}
\DeclarePairedDelimiter\ceil{\lceil}{\rceil}
\DeclarePairedDelimiter\floor{\lfloor}{\rfloor}

\SectionNumbersOn

\author{Guillaume Acke}
\author{Daria Van Hende}
\author{Ruben Van der Stichelen}
\author{Patrick Bultinck}
\email{patrick.bultinck@ugent.be}
\affiliation{Department of Chemistry, Ghent University, Krijgslaan 281 (S3), B-9000 Ghent, Belgium}

\title[qit-edit]
  {Can the electron density be interpreted information-theoretically? A critical analysis using quantum information theory}

\abbreviations{}
\keywords{}

\begin{document}

\begin{abstract}
  Many quantum chemical similarity measures have been derived and substantiated by applying concepts and quantities from information theory to the electron density. To justify the use of information theory, the electron density is usually equated to a probability distribution, despite the fact that such an assumption can give rise to inconsistencies such as negative Kullback-Leibler divergences. In this work we show, using quantum information theory, that both interpreting electron densities as probability distributions as well as any pragmatic normalization thereof gives rise to far-reaching information theoretical inconsistencies. By applying the theory of open quantum subsystems to quantum chemical partitions, we show using thought experiments that many statements concerning the information content of atoms in molecules need to be reconsidered. This study represents a step towards generalizing and reformulating the information theoretical approach to conceptual quantum chemistry beyond its classical perspective.
\end{abstract}

\section{Introduction}

A guiding principle in designing novel compounds is to search for chemicals that are `similar' to compounds with desirable properties \cite{bultinck2005a}. By staying within a certain `radius of similarity', it is assumed that desired properties can be preserved, while other beneficial properties such as solubility, etc. can be optimized \cite{bender2004a}. Many quantum chemical similarity measures have been proposed that aim to assess the similarity between molecules \cite{bultinck2005a}. Measures that are based on the electron density are usually preferred, under the premise that the Hohenberg-Kohn theorems establish that the electron density suffices to determine the Hamiltonian and ground state energy of a system \cite{parr1994a}. As such, \emph{in principle}, \emph{all} possible differences between two systems are completely determined by the differences in electron densities, although \emph{in practice} the required functionals to quantify \emph{all} those differences based on the underlying wavefunctions are unknown \cite{engel2011a}.

One of the key insights of Nalewajski and Parr \cite{nalewajski2000a} was that a similarity measure based on the electron density can also be used to define `atoms in the molecule': partition the electron density $\rho^{(1)}$ of a system in such a way that the densities of the atoms in the molecule $\rho^{(1)}_A$ are as similar as possible to their isolated counterparts $\rho^{(1)}_{A^0}$ (called `proatoms' \cite{hirshfeld1977a}). They showed that when choosing following similarity measure
\begin{equation}
  \sum\limits_{A} \int \rho^{(1)}_A(\vb{r}) \ln \left( \frac{\rho^{(1)}_A (\vb{r})}{\rho^{(1)}_{A^0}(\vb{r})} \right) \dd \vb{r}
\end{equation}
the optimal way to partition a molecule corresponds to the so-called `stockholder' partitioning of Hirshfeld \cite{hirshfeld1977a}, where the share of electron density attributed to an atom is determined according to the fraction of `proatomic' density in the corresponding promolecule (constructed from placing the proatoms at their respective positions in the molecular framework)
\begin{equation}
  \rho^{(1)}_A(\vb{r}) = \frac{\rho^{(1)}_{A^0}(\vb{r})}{\sum\limits_{B} \rho^{(1)}_{B^0}(\vb{r})} \rho^{(1)}(\vb{r}) \ .
\end{equation}

Unfortunately, Nalewajski and Parr tried to further substantiated this similarity measure by attempting to frame it within an information-theoretic perspective, ignoring -- in line with a deep-rooted confusion that persists in the conceptual quantum chemistry community to this day -- the difference between electron densities and probability distributions. Although a \emph{symptom} of this discrepancy, namely that the electron density integrates to the number of electrons and not one, was pragmatically resolved by normalizing the electron density to the shape function \cite{ayers2006a}, the difference between the resulting normalized shape function and proper probability distributions has not been addressed.

However, information theory requires proper probability distributions to correctly quantify the resources needed for optimal coding and error-free communication \cite{nielsen2010a} and any deviation from the axioms that underlie information theory implies that the estimated information content becomes operationally meaningless and that the proposed theory cannot be called an information theory. In this study we will show that the electron density is not a probability distribution and that electron density similarity measures are not compatible with information theory. We argue that differences in electron density should consistently be interpreted as differences in number densities, which have more to do with populations and charges than with probabilities. 

\section{Theory}

A probability space is a triple $(\Omega, \mathcal{F}, p)$, comprising \cite{grimmett2020a}
\begin{itemize}
  \item a sample space $\Omega$, which is the set of all possible outcomes from an experiment
  \item a $\sigma$-field $\mathcal{F}$ of subsets of $\Omega$
  \item a probability measure $p$ on $(\Omega, \mathcal{F})$
\end{itemize}
In the current context, the experiments associated with the sample space are those related with detecting electron at certain positions. 

In order to be able to theoretically describe those detections, we will first introduce the operators that describe the (un)occupation of spin-orbitals in \cref{sec:occupation-number-operators} and partitions (subsets) over those operators in \cref{sec:one-spin-orbital-reduced-density-matrix}. After porting those concepts to spin-position space using field operators in \cref{sec:field-operators}, we will show that the sample space of a position $\vb{r}$ is given by the outcomes: no electrons, one (and only one) up-spin electron, one (and only one) down-spin electron, and one-up and one-down spin electron in \cref{sec:fock-space-position}. This will allow us to show that the electron density is a number-of-electrons weighted sum of probabilities and not a probability distribution. In \cref{sec:shape-function} we will show that normalizing the electron density does not lead to a probability distribution. These findings will allow us to show in \cref{sec:electron-density-information-theoretical-measures} that measures based on the electron density are not information theoretical measures. After introducing a quantum information theory of atoms in molecules in \cref{sec:quantum-information-theory-atoms-in-molecules}, we will be able to show in \cref{sec:hirshfeld-I-information-theory} that the information-theoretic interpretation of Nalewajski and Parr is unfounded and can be abonded.

\subsection{Measuring spin-orbital (un)occupation number operators gives the probability of finding (holes)electrons}\label{sec:occupation-number-operators}

Given a basis of M orthonormal spin-orbitals $\left\{ \ket{\phi_P} \right\}$, we can construct a Fock space $\mathcal{F}$ consisting of the direct sum of Fock subspaces $\mathcal{F}(M,N)$ with a given number of particles $N$ \cite{helgaker2013a,acke2020a}
\begin{equation}
  \mathcal{F} = \mathcal{F}(M,0) \oplus \mathcal{F}(M,1) \oplus \ldots \oplus \mathcal{F}(M,M) \ .
\end{equation}
This Fock space $\mathcal{F}$ is spanned by occupation number vectors
\begin{equation}
  \ket{\vb{k}} = \ket{k_1, k_2, \ldots, k_M}, \quad k_P = \begin{cases} 1, & \text{if $\ket{\phi_P}$ occupied} \\ 0, & \text{if $\ket{\phi_P}$ unoccupied} \end{cases} \ ,
\end{equation}
which are eigenstates of the occupation-number and unoccupation-number operators $\hat{n}(P)$ and $\hat{h}(P)$
\begin{align}
  \hat{n}(P) \ket{\vb{k}} & = \delta_{k_P1} \ket{\vb{k}} \\
  \hat{h}(P) \ket{\vb{k}} & = \delta_{k_P0} \ket{\vb{k}} \ ,
\end{align}
where
\begin{equation}
  \hat{n}(P) = \hat{a}^{\dagger}_P \hat{a}_P 
\end{equation}
and
\begin{equation}
  \hat{h}(P) = \hat{a}_P \hat{a}^{\dagger}_P
\end{equation}
count the number of electrons/holes in spin orbital $\ket{\phi_P}$ respectively. Each subspace of $\mathcal{F}$ contains those occupation number vectors that have the same eigenvalue for the particle-number operator $\hat{N}$ and the hole-number operator $\hat{H}$ 
\begin{align}
  \hat{N} & = \sum\limits_{P=1}^{M} \hat{n}(P) = \sum\limits_{P=1}^{M} \hat{a}^{\dagger}_P \hat{a}_P \\
  \hat{H} & = \sum\limits_{P=1}^{M} \hat{h}(P) = \sum\limits_{P=1}^{M} \hat{a}_P \hat{a}^{\dagger}_P \ .
\end{align}

A given wavefunction $\ket{\Psi}$ with associated density matrix operator $\hat{\rho}$
\begin{equation}
  \hat{\rho} = \ket{\Psi} \bra{\Psi} 
\end{equation}
can be expanded in this occupation number vector basis
\begin{equation}
  \ket{\Psi} = \sum\limits_{\vb{k}} \ket{\vb{k}} \bra{\vb{k}} \ket{\Psi} = \sum\limits_{\vb{k}} c_{\vb{k}} \ket{\vb{k}}
\end{equation}
and is normalized to one
\begin{equation}
  \Tr_{\mathcal{F}}(\hat{\rho}) = \sum\limits_{\vb{k}} \braket{\vb{k}}{\Psi} \braket{\Psi}{\vb{k}} = \sum\limits_{\vb{k}} \left| c_{\vb{k}} \right|^2 = 1 \ .
\end{equation}
As Born's rule\cite{born1926a} states that the probability of finding a system in a state $\ket{\vb{k}}$ is proportional to the square of the amplitude of the wavefunction $c_{\vb{k}}$ at that state, the average of the occupation-number $\hat{n}(P)$ and unoccupation-number $\hat{h}(P)$ operators 
\begin{align}
  \mel{\Psi}{\hat{n}(P)}{\Psi} & = \sum\limits_{\vb{k}} \delta_{k_P1} \left| c_{\vb{k}} \right|^2 = p_{1,P} \\
  \mel{\Psi}{\hat{h}(P)}{\Psi} & = \sum\limits_{\vb{k}} \delta_{k_P0} \left| c_{\vb{k}} \right|^2 = p_{0,P}
\end{align}
are the \emph{probabilities} of measuring a particle/hole at spin-orbital $P$, as these are the sums of the probabilities $\left| c_{\vb{k}} \right|^2$ that the system is found in a state $\ket{\vb{k}}$ that has spin-orbital $\ket{\phi_P}$ occupied/unoccupied.

\subsection{The one-spin-orbital reduced density matrix gathers a probability distribution on its diagonal}\label{sec:one-spin-orbital-reduced-density-matrix}

Given a partition of spin-orbitals into subsets $A = \left\{ \ket{\phi_1}, \ldots, \ket{\phi_L} \right\}$ and $B = \left\{ \ket{\phi_{L+1}}, \ldots, \ket{\phi_M} \right\}$, we can obtain the reduced density operator for the subset $A$ by tracing over the Fock space $\mathcal{F}_B$ of $B$
\begin{equation}
  \Tr_{B} = \Tr_{\mathcal{F}_B} \left( \hat{\rho} \right) = \hat{\rho}_A \ ,
\end{equation}
with
\begin{equation}
  \mathcal{F}_B =  \mathcal{F}(|B|,0) \oplus \mathcal{F}(|B|,1) \oplus \ldots \oplus \mathcal{F}(|B|,|B|) \ ,
\end{equation}
where $|B|$ is the number of spin-orbitals in $B$. 

Equivalently, since $\mathcal{F} = \mathcal{F}_A \otimes \mathcal{F}_B$, we can construct operator representation $\hat{O}_A$ of the elements in $\mathcal{F}_A$ and determine the values of the corresponding matrix representation as average values over those operators
\begin{equation}
  \Tr_{\mathcal{F}}(\hat{O}_A \hat{\rho}) = \sum\limits_{\vb{k}} \braket{\vb{k}}{\hat{O}_A \Psi} \braket{\Psi}{\vb{k}} = \mel{\Psi}{\hat{O}_A}{\Psi}
\end{equation}
to obtained reduced density matrices $\vb*{\rho}_A$. For instance, if we take the subset $A$ to consist of one spin-orbital $\ket{\phi_P}$, the associated one-orbital reduced density operator $\hat{\rho}_P$ can be represented in $\mathcal{F}_P$ as the one-orbital reduced density matrix $\vb*{\rho}_P$
\begin{equation}
  \vb*{\rho}_P = \begin{bmatrix}
    \expval{\hat{a}_P \hat{a}^{\dagger}_P} & \expval{\hat{a}_P \hat{a}_P} \\
    \expval{\hat{a}^{\dagger}_P \hat{a}^{\dagger}_P} & \expval{\hat{a}^{\dagger}_P \hat{a}_P}
  \end{bmatrix} = \begin{bmatrix}
    \expval{\hat{h}(P)} & 0 \\
    0 & \expval{\hat{n}(P)} \
  \end{bmatrix} = \begin{bmatrix}
    p_{0,P} & 0 \\
    0 & p_{1,P} \
  \end{bmatrix} \ .
\end{equation}
As such, in the case of one spin-orbital, the associated one-spin-orbital reduced density matrix gathers the probabilities of measuring holes/particles at spin-orbital $P$ on its diagonal.

\subsection{Electron densities in spin-position space are averages over field operators}\label{sec:field-operators}

Before discussing the relation between electron densities and probability distributions, we will first express the theory derived above in the basis of position orbitals $\left\{ \ket{\vb{r}} \right\}$ as the electron density in quantum chemistry is predominantly studied in that basis. This is the basis of eigenvectors of the position operator $\hat{\vb{r}}$
\begin{equation}
  \hat{\vb{r}} \ket{\vb{r}} = \vb{r} \ket{\vb{r}} \ ,
\end{equation}
and we can construct a basis of spin-orbitals by explicitly taking spin into account
\begin{equation}
  \hat{\vb{x}} \ket{\vb{x}} = \vb{x} \ket{\vb{x}} \ ,
\end{equation}
with $\ket{\vb{x}} = \ket{\vb{r}\sigma}$, where $\sigma$ is a projected spin. The associated creation and annihilation operators are called field operators $\left\{ \hat{\psi}^{\dagger}_{\vb{x}}, \hat{\psi}_{\vb{x}} \right\}$ and they can be projected onto the above (generally incomplete) basis of spin-orbitals $\left\{ \ket{\phi}_P \right\}$ as 
\begin{equation}
  \text{proj}_{\left\{ \ket{\phi}_P \right\}} \hat{\psi}^{\dagger}_{\vb{x}} \ket{0} = \sum\limits_P \ket{\phi_P}\braket{\phi_P}{\vb{x}} = \sum\limits_P \ket{\phi_P} \phi^{*}_P(\vb{x}) = \sum\limits_{P} \phi^{*}_P(\vb{x}) \hat{a}^{\dagger}_P \ket{0} \ ,
\end{equation}
which we will denote as
\begin{align}  
  \hat{\psi}^{\dagger}_{\vb{x}} & \mapsto \sum\limits_{P} \hat{a}^{\dagger}_P \phi^{*}_P(\vb{x}) \\
  \hat{\psi}_{\vb{x}} & \mapsto \sum\limits_{P} \hat{a}_P \phi_P(\vb{x}) \ .
\end{align}

\subsection{The electron density is not a probability distribution but a number-of-electrons weighted sum of probabilities}\label{sec:fock-space-position}

A given position orbital $\ket{\vb{r}}$ can be spin-resolved into two spin components $\left\{ \ket{\vb{r}\uparrow}, \ket{\vb{r}\downarrow} \right\}$.
As this orbital basis is orthonormal, we can construct a Fock space $\mathcal{F}_{\ket{\vb{r}}}$ consisting of the occupation number vector basis $\left\{ \ket{0_{\vb{r}\uparrow}0_{\vb{r}\downarrow}}, \ket{1_{\vb{r}\uparrow}0_{\vb{r}\downarrow}}, \ket{0_{\vb{r}\uparrow}1_{\vb{r}\downarrow}}, \ket{1_{\vb{r}\uparrow}1_{\vb{r}\downarrow}} \right\}$. We will denote this basis using the respective shorthands $\left\{ \ket{00}_{\vb{r}}, \ket{10}_{\vb{r}}, \ket{01}_{\vb{r}}, \ket{11}_{\vb{r}} \right\}$. The reduced density operator associated with the Fock space $\mathcal{F}_{\ket{\vb{r}}}$ can be obtained by tracing over the Fock space associated with its complement $\mathcal{F}_{\overline{\ket{\vb{r}}}}$ with $\overline{\ket{\vb{r}}}$ the collection of all position orbitals \emph{except} $\ket{\vb{r}}$ 
\begin{equation}
  \hat{\rho}_{\ket{\vb{r}}} = \Tr_{\mathcal{F}_{\overline{\ket{\vb{r}}}}} \left( \hat{\rho} \right) \ .
\end{equation}
The trace over the environment can be expressed as averages over the field operators $\left\{ \hat{\psi}^{\dagger}_{\vb{r}\sigma}, \hat{\psi}_{\vb{r}\sigma} \right\}$ associated with the spin-position orbital $\ket{\vb{r}\sigma}$, leading to the one-position-orbital density matrix \cite{boguslawski2015a} expressed in the basis $\left\{ \ket{00}_{\vb{r}}, \ket{10}_{\vb{r}}, \ket{01}_{\vb{r}}, \ket{11}_{\vb{r}} \right\}$
\begin{equation}
  \vb*{\rho}_{\ket{\vb{r}}} = \begin{bmatrix}
    \expval{\hat{\psi}_{\vb{r}\uparrow} \hat{\psi}_{\vb{r}\downarrow} \hat{\psi}^\dagger_{\vb{r}\downarrow} \hat{\psi}^\dagger_{\vb{r}\uparrow}} & 
    \expval{\hat{\psi}_{\vb{r}\uparrow} \hat{\psi}_{\vb{r}\downarrow} \hat{\psi}^\dagger_{\vb{r}\downarrow} \hat{\psi}_{\vb{r}\uparrow}} & 
    \expval{\hat{\psi}_{\vb{r}\uparrow} \hat{\psi}_{\vb{r}\downarrow} \hat{\psi}_{\vb{r}\downarrow} \hat{\psi}^\dagger_{\vb{r}\uparrow}} & 
    \expval{\hat{\psi}_{\vb{r}\uparrow} \hat{\psi}_{\vb{r}\downarrow} \hat{\psi}_{\vb{r}\downarrow} \hat{\psi}_{\vb{r}\uparrow}} 
    \\
    \expval{\hat{\psi}^{\dagger}_{\vb{r}\uparrow} \hat{\psi}_{\vb{r}\downarrow} \hat{\psi}^\dagger_{\vb{r}\downarrow} \hat{\psi}^\dagger_{\vb{r}\uparrow}} & 
    \expval{\hat{\psi}^\dagger_{\vb{r}\uparrow} \hat{\psi}_{\vb{r}\downarrow} \hat{\psi}^\dagger_{\vb{r}\downarrow} \hat{\psi}_{\vb{r}\uparrow}} & 
    \expval{\hat{\psi}^{\dagger}_{\vb{r}\uparrow} \hat{\psi}_{\vb{r}\downarrow} \hat{\psi}_{\vb{r}\downarrow} \hat{\psi}^\dagger_{\vb{r}\uparrow}} & 
    \expval{\hat{\psi}^\dagger_{\vb{r}\uparrow} \hat{\psi}_{\vb{r}\downarrow} \hat{\psi}_{\vb{r}\downarrow} \hat{\psi}_{\vb{r}\uparrow}} 
    \\
    \expval{\hat{\psi}_{\vb{r}\uparrow} \hat{\psi}^{\dagger}_{\vb{r}\downarrow} \hat{\psi}^\dagger_{\vb{r}\downarrow} \hat{\psi}^\dagger_{\vb{r}\uparrow}} & 
    \expval{\hat{\psi}_{\vb{r}\uparrow} \hat{\psi}^{\dagger}_{\vb{r}\downarrow} \hat{\psi}^\dagger_{\vb{r}\downarrow} \hat{\psi}_{\vb{r}\uparrow}} & 
    \expval{\hat{\psi}_{\vb{r}\uparrow} \hat{\psi}^\dagger_{\vb{r}\downarrow} \hat{\psi}_{\vb{r}\downarrow} \hat{\psi}^\dagger_{\vb{r}\uparrow}} & 
    \expval{\hat{\psi}_{\vb{r}\uparrow} \hat{\psi}^\dagger_{\vb{r}\downarrow} \hat{\psi}_{\vb{r}\downarrow} \hat{\psi}_{\vb{r}\uparrow}} 
    \\
    \expval{\hat{\psi}^{\dagger}_{\vb{r}\uparrow} \hat{\psi}^{\dagger}_{\vb{r}\downarrow} \hat{\psi}^\dagger_{\vb{r}\downarrow} \hat{\psi}^\dagger_{\vb{r}\uparrow}} & 
    \expval{\hat{\psi}^\dagger_{\vb{r}\uparrow} \hat{\psi}^{\dagger}_{\vb{r}\downarrow} \hat{\psi}^\dagger_{\vb{r}\downarrow} \hat{\psi}_{\vb{r}\uparrow}} & 
    \expval{\hat{\psi}^{\dagger}_{\vb{r}\uparrow} \hat{\psi}^\dagger_{\vb{r}\downarrow} \hat{\psi}_{\vb{r}\downarrow} \hat{\psi}^\dagger_{\vb{r}\uparrow}} & 
    \expval{\hat{\psi}^\dagger_{\vb{r}\uparrow} \hat{\psi}^\dagger_{\vb{r}\downarrow} \hat{\psi}_{\vb{r}\downarrow} \hat{\psi}_{\vb{r}\uparrow}} 
  \end{bmatrix} \ .
  \label{eq:one-position-orbital-density-matrix}
\end{equation}
For a pure state $\ket{\Psi}$ with fixed particle-number and spin symmetry $\vb*{\rho}_{\ket{\vb{r}}}$ will be diagonal
\begin{equation}
  \vb*{\rho}_{\ket{\vb{r}}} = \begin{bmatrix}
    \expval{\hat{\psi}_{\vb{r}\uparrow} \hat{\psi}_{\vb{r}\downarrow} \hat{\psi}^\dagger_{\vb{r}\downarrow} \hat{\psi}^\dagger_{\vb{r}\uparrow}} & 0 & 0 & 0
    \\
    0 & 
    \expval{\hat{\psi}^\dagger_{\vb{r}\uparrow} \hat{\psi}_{\vb{r}\downarrow} \hat{\psi}^\dagger_{\vb{r}\downarrow} \hat{\psi}_{\vb{r}\uparrow}} & 0 & 0
    \\
    0 & 0 & 
    \expval{\hat{\psi}_{\vb{r}\uparrow} \hat{\psi}^\dagger_{\vb{r}\downarrow} \hat{\psi}_{\vb{r}\downarrow} \hat{\psi}^\dagger_{\vb{r}\uparrow}} & 0
    \\
    0 & 0 & 0 &
    \expval{\hat{\psi}^\dagger_{\vb{r}\uparrow} \hat{\psi}^\dagger_{\vb{r}\downarrow} \hat{\psi}_{\vb{r}\downarrow} \hat{\psi}_{\vb{r}\uparrow}} 
  \end{bmatrix} \ ,
\end{equation}
as the trace over the environment cannot couple changing particle-number or spin within the subsystem $\ket{\vb{r}}$ \cite{boguslawski2013a, boguslawski2015a}. This implies that $\vb*{\rho}_{\ket{\vb{r}}}$ can be rewritten in terms of the particle-number and spin conserving occupation-number $\hat{n}(\vb{r})$ and unoccupation-number $\hat{h}(\vb{r})$ operators 
\begin{align}
  \hat{n}(\vb{r}) & = \sum\limits_{\sigma=\uparrow,\downarrow} \hat{n}(\vb{r}\sigma) = \sum\limits_{\sigma=\uparrow,\downarrow} \hat{\psi}^{\dagger}_{\vb{r}\sigma} \hat{\psi}_{\vb{r}\sigma} \\
  \hat{h}(\vb{r}) & = \sum\limits_{\sigma=\uparrow,\downarrow} \hat{h}(\vb{r}\sigma) = \sum\limits_{\sigma=\uparrow,\downarrow} \hat{\psi}_{\vb{r}\sigma} \hat{\psi}^{\dagger}_{\vb{r}\sigma}
\end{align}
as
\begin{equation}
  \vb*{\rho}_{\ket{\vb{r}}} = \begin{bmatrix}
    \expval{\hat{h}(\vb{r}\uparrow) \hat{h}(\vb{r}\downarrow)} & 0 & 0 & 0
    \\
    0 & 
    \expval{\hat{n}(\vb{r}\uparrow) \hat{h}(\vb{r}\downarrow)} & 0 & 0
    \\
    0 & 0 & 
    \expval{\hat{h}(\vb{r}\uparrow) \hat{n}(\vb{r}\downarrow)} & 0
    \\
    0 & 0 & 0 &
    \expval{\hat{n}(\vb{r}\uparrow) \hat{n}(\vb{r}\downarrow)}
  \end{bmatrix} \ ,
\end{equation}
where we have used the fact that field operators corresponding to different spins anticommute 
\begin{equation}
  \lbrack \hat{\psi}^{\dagger}_{\vb{r}\sigma}, \hat{\psi}_{\vb{r}\sigma'} \rbrack_+ = 0, \quad \forall \sigma \neq \sigma' \ .
\end{equation}

We will denote this one-position orbital density matrix as
\begin{equation}
  \vb*{\rho}_{\ket{\vb{r}}} = \begin{bmatrix}
    p_{00,\vb{r}} & 0 & 0 & 0 \\
    0 & p_{10,\vb{r}} & 0 & 0 \\
    0 & 0 & p_{01, \vb{r}} & 0 \\
    0 & 0 & 0 & p_{11,\vb{r}} 
  \end{bmatrix} \ ,
\end{equation}
where $p_{00, \vb{r}}, p_{10,\vb{r}}, p_{01,\vb{r}}$ and $p_{11,\vb{r}}$ are the probabilities that the position orbital $\ket{\vb{r}}$ is occupied by no, one up, one down and one up and one down electrons. These probabilities are normalized 
\begin{equation}
  1 = p_{00,\vb{r}} + p_{10,\vb{r}} + p_{01,\vb{r}} + p_{11,\vb{r}} \ ,
\end{equation}
and the events that underlie those probabilities are mutually exclusive and form a sample space $\Omega$. As such $p_{X,\vb{r}}$ is a genuine probability distribution with the events to find no, one up, one down or one up and one down electron at the position orbital $\ket{\vb{r}}$.

In contrast, the electron density $\rho^{(1)}$ at position orbital $\ket{\vb{r}}$ is given by the average of the occupation-number operator $\hat{n}$
\begin{equation}
  \rho^{(1)}(\vb{r}) = \expval{\hat{n}(\bm{r})} = \expval{\hat{\psi}^{\dagger}_{\vb{r}} \hat{\psi}_{\vb{r}}} = \sum\limits_{\sigma=\uparrow,\downarrow} \expval{\hat{\psi}^{\dagger}_{\vb{r}\sigma} \hat{\psi}_{\vb{r}\sigma}} = \sum\limits_{\sigma=\uparrow,\downarrow} \expval{\hat{n}(\vb{r}\sigma)} \ .
  \label{eq:electron-density}
\end{equation}
As the anticommutator of a creation and annihilation field operator associated with a given spin-position orbital $\ket{\vb{r}\sigma}$ is given by
\begin{equation}
  1 = \lbrack \hat{\psi}_{\vb{r}\sigma}, \hat{\psi}^{\dagger}_{\vb{r}\sigma} \rbrack_{+} = \hat{\psi}_{\vb{r}\sigma} \hat{\psi}^{\dagger}_{\vb{r}\sigma} + \hat{\psi}^{\dagger}_{\vb{r}\sigma} \hat{\psi}_{\vb{r}\sigma} = \hat{h}(\vb{r}\sigma) + \hat{n}(\vb{r}\sigma) \ ,
\end{equation}
we can insert this commutator into the expression for the electron density (see \cref{eq:electron-density}) , where we set the spin of the inserted anticommutator opposite to the spin inside the summation of \cref{eq:electron-density} \cite{karafiloglou1988a, karafiloglou1990a, parrondo1994a, papanikolaou2008a, karafiloglou2009a, kyriakidou2017a}
\begin{align}
  \rho^{(1)}(\vb{r}) & = \sum\limits_{\sigma=\uparrow,\downarrow} \expval{\hat{n}(\vb{r}\sigma) \left( \hat{n}(\vb{r}\sigma') + \hat{h}(\vb{r}\sigma') \right)} \\
  & = \sum\limits_{\sigma=\uparrow,\downarrow} \expval{\hat{n}(\vb{r}\sigma) \hat{n}(\vb{r}\sigma')} + \sum\limits_{\sigma=\uparrow,\downarrow} \expval{\hat{n}(\vb{r}\sigma) \hat{h}(\vb{r}\sigma')}  \ ,
\end{align}
with $\sigma' \neq \sigma$. Due to the anticommutation rules of fermionic field operators, this reduces to
\begin{align}
  \rho^{(1)}(\vb{r}) & = 2 \expval{\hat{n}(\vb{r}\uparrow) \hat{n}(\vb{r}\downarrow)} + \expval{\hat{n}(\vb{r}\uparrow) \hat{h}(\vb{r}\downarrow)} + \expval{\hat{h}(\vb{r}\uparrow) \hat{n}(\vb{r}\downarrow)} \nonumber \\
  & = 2 p_{11,\vb{r}} + p_{10,\vb{r}} + p_{01,\vb{r}} \ .
  \label{eq:electron-density-in-terms-of-probabilities}
\end{align}
As such, the electron density is \emph{not} the probability to find an electron at position $\vb{r}$, but is a \emph{number-of-electrons weighted sum} of the probabilities that position orbital $\ket{\vb{r}}$ is occupied by no, one or two electrons (i.e. a number density).

We can reinforce this finding by doing the above analysis for the hole density $\eta^{(1)}$
\begin{equation}
  \eta^{(1)}(\vb{r}) = \expval{\hat{h}(\vb{r})} = \expval{\hat{\psi}_{\vb{r}} \hat{\psi}^{\dagger}_{\vb{r}}} = \sum\limits_{\sigma=\uparrow,\downarrow} \expval{\hat{\psi}_{\vb{r}\sigma} \hat{\psi}^{\dagger}_{\vb{r}\sigma}} = \sum\limits_{\sigma=\uparrow,\downarrow} \expval{\hat{h}(\vb{r}\sigma)}\ ,
\end{equation}
which leads to 
\begin{align}
  \eta^{(1)}(\vb{r}) & = 2 \expval{\hat{h}(\vb{r}\uparrow) \hat{h}(\vb{r}\downarrow)} + \expval{\hat{h}(\vb{r}\uparrow) \hat{n}(\vb{r}\downarrow)} + \expval{\hat{n}(\vb{r}\uparrow) \hat{h}(\vb{r}\downarrow)} \nonumber \\
  & = 2 p_{00,\vb{r}} + p_{01,\vb{r}} + p_{10,\vb{r}} \ .
\end{align}
As such, the sum of the electron density $\rho^{(1)}$ and the hole density $\eta^{(1)}$ at a position orbital $\ket{\vb{r}}$ is equal to
\begin{equation}
  \rho^{(1)}(\vb{r}) + \eta^{(1)}(\vb{r}) = 2 (p_{00,\vb{r}} + p_{10,\vb{r}} + p_{01,\vb{r}} + p_{11,\vb{r}}) = 2 \ .
\end{equation}
As the electron density is thus bounded by two (i.e. if a position orbital is occupied by two electrons), it cannot be interpreted or used as a probability. Rather, it is the average number of electrons at that position orbital $\ket{\vb{r}}$ and it always sums with the average number of holes at that position orbital to two.

\subsection{The spin-resolved electron/hole density}

In contrast, if we do the above analysis for a spin-position-orbital $\ket{\vb{r}\sigma}$, we obtain the one-spin-position-orbital density matrix in the occupation number basis $\left\{ \ket{0_{\vb{r}\sigma}}, \ket{1_{\vb{r}\sigma}} \right\}$
\begin{equation}
  \vb*{\rho}_{\ket{\vb{r}\sigma}} = \begin{bmatrix}
    \expval{\hat{h}(\vb{r}\sigma)} & 0 \\
    0 & \expval{\hat{n}(\vb{r}\sigma)}
  \end{bmatrix} \ ,
\end{equation}
which we denote as 
\begin{equation}
  \vb*{\rho}_{\ket{\vb{r}\sigma}} = \begin{bmatrix}
    p_{0,\vb{r}\sigma} & 0 \\
    0 & p_{1,\vb{r}\sigma}
  \end{bmatrix} \ .
\end{equation}
In this case, the associated spin-resolved electron density $\rho^{(1)}(\vb{r}\sigma)$ is given by 
\begin{equation}
  \rho^{(1)}(\vb{r}\sigma) = \expval{\hat{n}(\vb{r}\sigma)} = p_{1,\vb{r}\sigma} \ ,
\end{equation}
and is equal to the probability that $\ket{\vb{r}\sigma}$ is occupied.
As such, a probabilistic interpretation of a density \emph{can} hold true if there are only two possible occupations: empty or occupied.
The resulting probabilities \emph{are} normalized $p_{0,\vb{r}\sigma} + p_{1,\vb{r}\sigma} = 1$, where the mutually exclusive events are that the spin-orbital $\ket{\vb{r}\sigma}$ is either occupied or unoccupied.

In contrast, the electron density is specified in terms of a position orbital, which has \emph{four} possible occupations. As the electron density is a single number-weighted sum of the associated probabilities, we cannot obtain those underlying probabilities directly from the electron density alone (although \emph{in principle} there should exist some functional that describes that mapping).

\subsection{Normalizing the electron density to the shape function does not lead to a probability distribution}\label{sec:shape-function}

As the particle number operator $\hat{N}$ is given by an integral over the position occupation-number operator $\hat{n}(\vb{r})$
\begin{equation}
  \hat{N} = \int \hat{n}(\vb{r}) \dd \vb{r} \ ,
\end{equation} 
the electron density integrates to the number of electrons $N$ 
\begin{equation}
  \int \rho^{(1)}(\vb{r}) \dd \vb{r} = N \ .
\end{equation}
Given the number of electrons, the electron density can be normalized to the shape function $\sigma^{(1)}$ \cite{parr1983a}
\begin{equation}
  \sigma^{(1)}(\vb{r}) = \frac{\rho^{(1)}(\vb{r})}{N} \ .
\end{equation}
However, normalizing the electron density $\rho^{(1)}$ with the total number of electrons $N$ to the shape function $\sigma^{(1)}$ does not lead to a probability distribution, as
\begin{equation}
  \sigma^{(1)}(\vb{r}) = \frac{\rho^{(1)}(\vb{r})}  {N} = \frac{2 p_{11, \vb{r}} + p_{10,\vb{r}} + p_{01,\vb{r}}}{N} 
\end{equation}
still has no clear probabilistic interpretation. Normalizing the electron density \emph{mitigates} one of the symptoms of the electron density not being a probability distribution, but does not \emph{establish} a probability distribution, as there is no underlying sample space. Furthermore, normalizing in the case of the spin-resolved density $\rho^{(1)}(\vb{r}\sigma)$, replaces the probabilistic interpretation detailed above by 
\begin{equation}
  \sigma^{(1)}(\vb{r}\sigma) = \frac{p_{1,\vb{r}\sigma}}{N} \ ,
\end{equation}
which, ironically, destroys the normalization of that probability distribution.

\subsection{Measures based on the electron density are not (quantum) information theoretical measures}\label{sec:electron-density-information-theoretical-measures}

If we have access to a reduced density operator $\hat{\rho}_{\ket{\vb{r}}}$, then the associated von Neumann entropy is given by \cite{preskill1998a, nielsen2010a, wilde2013a, zeng2019a}
\begin{align}
  S(\hat{\rho}_{\ket{\vb{r}}}) & = - \Tr_{\mathcal{F}_{\ket{\vb{r}}}} \left( \hat{\rho}_{\ket{\vb{r}}} \ln \hat{\rho}_{\ket{\vb{r}}}\right) \\
  & = - \sum\limits_{i \in \left\{00, 10, 01, 11 \right\}} p_{i,\vb{r}} \ln(p_{i,\vb{r}}) \ ,
  \label{eq:von-neumann-entropy}
\end{align}
and quantifies the entanglement of the position orbital $\ket{\vb{r}}$ with the environment consisting of all position orbitals \emph{except} $\ket{\vb{r}}$ \cite{boguslawski2012b, boguslawski2015a}. We explicitly note that the trace is taken over the Fock space $\mathcal{F}_{\ket{\vb{r}}}$ associated with the position orbital $\ket{\vb{r}}$. The quantum relative entropy of another density operator $\hat{\tau}_{\ket{\vb{r}}}$ at position orbital $\ket{\vb{r}}$ is then given by
\begin{align}
  S(\hat{\rho}_{\ket{\vb{r}}} || \hat{\tau}_{\ket{\vb{r}}}) & = \Tr \left( \hat{\rho}_{\ket{\vb{r}}} \log \hat{\rho}_{\ket{\vb{r}}} \right) - \Tr \left( \hat{\rho}_{\ket{\vb{r}}} \log \hat{\tau}_{\ket{\vb{r}}} \right) \nonumber \\
  & = \sum\limits_{i \in \left\{00, 10, 01, 11 \right\}} p^{\vb*{\rho}}_{i,\vb{r}} \ln \left( \frac{p^{\vb*{\rho}}_{i,\vb{r}}}{p^{\vb*{\tau}}_{i,\vb{r}}} \right) \ ,
\end{align}
as $\vb*{\rho}_{\ket{\vb{r}}}$ and $\vb*{\tau}_{\ket{\vb{r}}}$ are simultanously diagonal in $\mathcal{F}_{\ket{\vb{r}}}$.

In contrast, the measure proposed by Nalewajski and Parr \cite{nalewajski2000a} is based on integrating  the electron density over the entire position space
\begin{equation}
  - \int \rho^{(1)}(\vb{r}) \ln \rho^{(1)}(\vb{r}) \dd \vb{r} \ .
\end{equation}
If we substitute in the derived expressions in terms of probabilities (see \cref{eq:electron-density-in-terms-of-probabilities})
\begin{equation}
  - \int \left( 2 p_{11,\vb{r}} + p_{10,\vb{r}} + p_{01,\vb{r}} \right) \ln \left( 2 p_{11,\vb{r}} + p_{10,\vb{r}} + p_{01,\vb{r}} \right) \dd \vb{r} \ ,
\end{equation}
we see that we are integrating over number-weighted probabilities associated with different sample spaces.  The random variable associated with position $\vb{r}$ with values $(0, \uparrow, \downarrow, \uparrow \downarrow)$ is \emph{different} from the random variable associated with position $\vb{r'}$ with its values $(0, \uparrow, \downarrow, \uparrow \downarrow)$. In other words, $p_{10,\vb{r}} = 1$ does \emph{not} imply that $p_{10,\vb{r'}} = 0$, with $\vb{r} \neq \vb{r'}$. As such, the proposal of Nalewajski and Parr \cite{nalewajski2000a} to relate their measure to an information theoretical measure seems to be based on a \emph{misinterpretation} of the entire position space as one random variable, which is a direct consequence of (mis)interpreting electron densities as probability distributions.

\subsection{Quantum information theories of atoms in molecules}\label{sec:quantum-information-theory-atoms-in-molecules}

As the original intent of introducing measures inspired by information theory was aimed at deriving partitioning schemes for atoms in molecules \cite{nalewajski2000a}, we show in this section what properties are required for coherent quantum information theories of atoms in molecules \cite{tubman2014a, francisco2023a, vanhende2024a}. This will allow us to show that electron density based similarity measures are to be interpreted in terms of charges/populations and not probabilities.

\subsubsection{Atomic weights lead to domain occupation number operators}

Atoms derived from the electron density $\rho^{(1)}$ lead to domains $\left\{ \rho^{(1)}_A(\vb{r}) \right\}$  that can be written in terms of weight functions $\left\{ w_A(\vb{r}) \right\}$ in position space
\begin{equation}
  \rho^{(1)}(\vb{r}) = \sum\limits_{A}^{N_{\text{atoms}}} w_A(\vb{r}) \rho^{(1)}(\vb{r}) = \sum\limits_{A}^{N_{\text{atoms}}} \rho^{(1)}_A(\vb{r}) \ ,
\end{equation}
with
\begin{equation}
  \sum\limits_{A}^{N_{\text{atoms}}} w_A(\vb{r}) = 1 \ ,
\end{equation}
and
\begin{equation}
  \forall A: 0 \leq w_A(\vb{r}) \leq 1 \ .
\end{equation}
In the case of Hirshfeld's stockholder partitioning \cite{hirshfeld1977a}, the weight function is constructed from a collection of proatomic densities $\left\{ \rho^{(1)}_{A^0} \right\}$, which are pasted onto their respective nuclear positions
\begin{equation}
  w_A(\vb{r}) = \frac{\rho^{(1)}_{A^0}(\vb{r})}{\sum\limits_{B} \rho^{(1)}_{B^0}(\vb{r})} \ .
\end{equation}
In the case of the Quantum Theory of Atoms in Molecules (QTAIM) \cite{bader1990a, bader1991a}, an atom in the molecule $A$ is considered bounded by a zero-flux surface $\partial A$
\begin{equation}
  \nabla \rho^{(1)}(\vb{r}) \cdot \vb{n_{A}}(\vb{r}) = 0 \ ,
\end{equation}
with $\vb{n_{A}}$ is the exterior normal vector at each position $\vb{r}$ of the surface of $\partial A$, leading to weight functions of the form \cite{ayers2015a}
\begin{align}
  w_{A}(\vb{r}) = \begin{cases}
    1, & \text{if}\ \vb{r} \in A \\
    0, & \text{if}\ \vb{r} \notin A
  \end{cases} \ .
\end{align}

The resulting weight function $w_{A}$ can then be used to construct domain occupation number operators on the atom $A$ and its environment $\overline{A}$ \cite{acke2020a}
\begin{align}
  \widehat{n}_{A} & = \int\limits w_{A}(\vb{r}) \hat{n}(\vb{r}) \dd \vb{r} \\
  \widehat{n}_{\overline{A}} & = \int\limits (1 - w_{A}(\vb{r})) \hat{n}(\vb{r}) \dd \vb{r} \ ,
\end{align}
which can be projected onto a given spin-orbital basis $\left\{ \ket{\phi_P} \right\}$ as
\begin{align}
  \hat{n}_A & = \int w_A(\vb{r}) \hat{n}(\vb{r}) \dd \vb{r} = \sum\limits_{\sigma} \int w_A(\vb{r}) \hat{n}(\vb{r}\sigma) \dd \vb{r} \\
  & \mapsto \sum\limits_{PQ} \left( \int w_A(\vb{r}) \phi_P^{*}(\vb{r}) \phi_Q(\vb{r}) \dd \vb{r} \right) \hat{a}^{\dagger}_P \hat{a}_Q = \sum\limits_{PQ} \Sigma^A_{PQ} \hat{a}^{\dagger}_P \hat{a}_Q \ ,
\end{align}
where $\Sigma^A_{PQ}$ is the atomic spin-orbital overlap matrix.

\subsubsection{Hirshfeld(-I) does not require an information theoretic context}\label{sec:hirshfeld-I-information-theory}

As posited by Heidar-Zadeh and coworkers \cite{heidar-zadeh2018a}, atoms in molecules should be chosen to resemble their isolated atoms (proatom) to the greatest possible extent. This implies that we quantify and minimize the dissimilarity $D$ between an atom in the molecule and its proatom. If we \emph{postulate} that this dissimilarity can be captured by the electron density, we obtain the following measure
\begin{equation}
  D_{\text{AIM-mol}, \text{pro-mol}} = \sum\limits_{A}^{N_{\text{atoms}}} D \left[ \rho^{(1)}_{A}, \rho^{(1)}_{A^0} \right] \ ,
\end{equation}
where a AIM-molecule is defined as the collection of atoms in molecules ($\text{AIM-mol} = \left\{ A \right\}$) and a promolecule is defined as the collection of proatoms ($\text{pro-mol} = \left\{ A^0 \right\}$).

In this context, Hirshfeld can be formulated in terms of a minimization of a Lagrangian $\Lambda$, where an electron density based similarity measure is constrained to variations that sum to a given molecular density $\rho^{(1)}$
\begin{align}
  \Lambda & \left[ \left\{ \rho^{(1)}_A \right\}; \left\{ \rho^{(1)}_{A^0} \right\}, \rho^{(1)} \right] = \nonumber \\
  & \quad \quad \sum\limits_{A} \int \rho^{(1)}_A(\vb{r}) \ln \left( \frac{\rho^{(1)}_A (\vb{r})}{\rho^{(1)}_{A^0}(\vb{r})} \right) \dd \vb{r} \nonumber \\
  & \quad \quad + \int \lambda(\vb{r}) \left( \rho^{(1)}(\vb{r}) - \sum\limits_{A} \rho^{(1)}_A (\vb{r}) \right) \dd \vb{r} \ ,
\end{align}
where variations in $\rho^{(1)}_A$ correspond to changes in the weights $w_A(\vb{r})$ in the domain occupation-number operator 
\begin{equation}
  \hat{n}(A) = \int w_{A}(\vb{r}) \hat{n}(\vb{r}) \dd \vb{r} \ .
\end{equation}
Minimizing this functional leads to the Hirshfeld or Stockholder partitioning (see Appendix of ref\cite{heidar-zadeh2018a}), where the electron density associated with the atom in the molecule $A$ is given by
\begin{equation}
  \rho^{(1)}_A(\vb{r}) = \frac{\rho^{(1)}_{A^0}(\vb{r})}{\sum\limits_{B} \rho^{(1)}_{B^0}(\vb{r})} \rho^{(1)}(\vb{r}) \ ,
\end{equation}
with associated weight 
\begin{equation}
  w_{A}(\vb{r}) = \frac{\rho^{(1)}_{A^0}(\vb{r})}{\sum\limits_{B} \rho^{(1)}_{B^0}(\vb{r})} \ .
\end{equation}
In Hirshfeld, the choice of charge (and state) of the pro-atoms is essentially arbitrary but remains fixed after a choice has been made. E.g., for \ce{LiF} one can choose ground state \ce{Li^0} and \ce{F^0} pro-atomic densities, but an equally valid choice would be \ce{Li^+} and \ce{F^-}.

This ambiguity is circumvented in Hirshfeld-I \cite{bultinck2007a}, where the proatoms are not considered isolated, but are put in a bath of electrons. In an iterative way, two Lagragians $\Lambda$ and $\Xi$ are minimized in turns until self-consistency is reached. In the Lagrangian $\Xi$, the proatoms are allowed to exchange electrons with an electron bath but are constrained to number-conserving variations of $\rho^{(1)}_{A^0}(\vb{r})$ 
\begin{align}
  \Xi & \left[ \left\{ \rho^{(1)}_{A^0} \right\}; \left\{ \rho^{(1)}_{A} \right\}, \rho^{(1)} \right] = \nonumber \\
  & \quad \quad \sum\limits_{A} \left( F[\rho^{(1)}_{A^0}] + \int\limits \rho^{(1)}_{A^0}(\vb{r}) v^A(\vb{r}) \dd \vb{r} \right) \nonumber \\
  & \quad \quad + \sum\limits_A \mu_A \int \left( \rho^{(1)}_{A^0}(\vb{r}) - \rho^{(1)}_{A}(\vb{r}) \right) \dd \vb{r}
\end{align}
where $F$ is the universal variational functional and $\nu(\vb{r})$ is the external potential \cite{perdew1982a}. Variations in $\rho^{(1)}_{A^0}$ lead to \cite{perdew1982a}
\begin{align}
  \rho^{(1)}_{A^0} & = \rho^{(1)}_{A^0}(\floor{N_A}) \left( \ceil{N_A} - N_A \right) \nonumber \\
  & \quad \quad + \rho^{(1)}_{A^0}(\ceil{N_A}) \left( N_A - \floor{N_A} \right) \ ,
\end{align}
where $\floor{N_A}$ and $\ceil{N_A}$ are used to denote the greatest integer less than or equal to $N_A$ and the smallest integer greater than or equal to $N_A$ respectively.

As such, Hirshfeld and Hirshfeld-I can both be derived \emph{solely} from statements about the number of electrons and the underlying number density. The similarity measure used \emph{cannot} be called the Kullback-Leibler divergence, as this statistical distance is only defined for two probability distributions, and those probability distributions consist of a sample space and the probabilities associated with that sample space. Even if we were to normalize the electron density to one, there still is no corresponding sample space to justify the use of information theoretic interpretations.

\subsubsection{Atomic densities obtained from stockholder partitions contain contributions from other atoms}

Stockholder partitions are usually classified as `fuzzy', because a position orbital is `shared' by several atoms, while the density $\rho^{(1)}_A$ assigned to an atom $A$ is considered as integral part of that atom. However, there are further conceptual consequences associated with the `fuzzy' nature of the stockholder partitioning. If we expand an atomic density $\rho^{(1)}_A$ in its underlying probabilities
\begin{align}
  \rho^{(1)}_A(\vb{r}) & = \sum\limits_{\sigma} \rho^{(1)}_A(\vb{r}\sigma) = \sum\limits_{\sigma} \expval{\hat{n}_A(\vb{r}\sigma) \left( \sum\limits_{B} \left( \hat{n}_B(\vb{r} \sigma') + \hat{h}_B(\vb{r} \sigma') \right) \right)} \nonumber \\
  & = \sum\limits_{\sigma} \left( \sum\limits_B \expval{\hat{n}_A(\vb{r}\sigma) \hat{n}_B(\vb{r} \sigma')} + \sum\limits_B \expval{\hat{n}_A(\vb{r}\sigma) \hat{h}_B(\vb{r} \sigma')} \right) \nonumber \\
  & = \sum\limits_B \left( 2 * p_{11,\vb{r},AB} + p_{10,\vb{r},AB} + p_{01,\vb{r},AB} \right) \ ,
\end{align}
we see that the density of atom $A$ is determined by probabilities which depend on \emph{all} other atoms. In contrast, in the case of so-called `binary' partitions such as QTAIM
\begin{align}
  \rho^{(1)}_A(\vb{r}) & = \sum\limits_{\sigma} \rho^{(1)}_A(\vb{r}\sigma) = \sum\limits_{\sigma} \expval{\hat{n}_A(\vb{r}\sigma) \left( \sum\limits_{B} \left( \hat{n}_B(\vb{r} \sigma') + \hat{h}_B(\vb{r} \sigma') \right) \right)} \nonumber \\
  & = \sum\limits_{\sigma} \left( \sum\limits_B \expval{\hat{n}_A(\vb{r}\sigma) \hat{n}_B(\vb{r} \sigma')} + \sum\limits_B \expval{\hat{n}_A(\vb{r}\sigma) \hat{h}_B(\vb{r} \sigma')} \right) \nonumber \\
  & = 2 * p_{11,\vb{r},AA} + p_{10,\vb{r},AA} + p_{01,\vb{r},AA}
\end{align}
these probabilities only depend on the occupations of atom $A$ itself. Consequently, the fuzzy nature of stockholder partitions extends much further than simply sharing the occupancies of a particular orbital position. As this has severe consequences when attempting to build a quantum information theory of atoms in molecules, we will focus in the following on QTAIM (`binary') partitions.

\subsubsection{Atomic density matrices can be obtained for single Slater determinants in the case of QTAIM}

As mentioned above, a quantum informational theory of atoms in molecules requires that we have access to the density matrix of the atom. In this section, we will show that such a density matrix can be obtained explicitly in the case of QTAIM partitions based on electron densities that originate from single Slater determinant wavefunction. Our analysis will be based on findings in the field of condensed matter physics and quantum information theory \cite{klich2006a, peschel2012a, song2012a} and provide a comprehensive framework for the concepts used by Tubman \cite{tubman2014a} and Pendas \cite{pendas2018a}, among others.

In the case of a single Slater determinant $\ket{\Phi}$
\begin{equation}
  \ket{\Phi} = \prod_{I=1}^{N} \widehat{a}^{\dagger}_I \ket{0} \ ,
\end{equation}
with $\ket{0}$ the physical vacuum, the physical wave function does not change when the occupied orbitals are rotated among each other. As such, we can focus on the domain spin-orbital overlap matrix of the occupied orbitals and diagonalize the resulting matrix $\Sigma^{\text{occ},A}$ (with $\dim(\Sigma^{\text{occ}, A}) = N$) 
\begin{equation}
  \Sigma^{\text{occ},A} = V^{A} d^{A} (V^{A})^{\dagger} \ ,
\end{equation}
where $V^{A}$ diagonalizes both $\Sigma^{\text{occ}, A}$ and $(1 - \Sigma^{\text{occ}, A}) = \Sigma^{\text{occ}, \overline{A}}$. If we rotate the spin-orbitals $\ket{\phi_P}$ according to $V^{A}$
\begin{equation}
  \ket{\tilde{\phi}_{I}} = \sum\limits_{I=1}^{N} (V^{A,\dagger})_{IP} \ket{\phi_P} \ ,
\end{equation}
we obtain the following representation of the single Slater determinant $\ket{\Phi}$
\begin{equation}
  \ket{\Phi} = \prod_{I=1}^{N} \hat{\tilde{a}}_{I} \ket{0} \ ,
\end{equation}
in the basis of the so-called domain naturals $\ket{\tilde{\phi}_{I}}$.

We can now construct two sets of $N$ orthonormal spin-orbitals, where one set is the set of domain spin-orbitals and the other set is the set of environment spin-orbitals \cite{klich2006a} (which is equivalent to the fragment/bath construction in Density Matrix Embedding Theory (DMET) in the case of a low-level Hartree-Fock wavefunction \cite{bulik2014a, wouters2016a})
\begin{align}
  \ket{\tilde{\phi}^{A}_{I}} & = \frac{\widehat{n}_{A}}{\sqrt{d^A_I}} \ket{\tilde{\phi}_{I}} \\
  \ket{\tilde{\phi}^{\overline{A}}_{I}} & = \frac{\widehat{n}_{\overline{A}}}{\sqrt{1 - d^A_I}} \ket{\tilde{\phi}_{I}} \ ,  
\end{align}
which leads to the orbitals with the following properties (see \cref{sec:orthogonality-orbitals-atoms})
\begin{align}
  \braket{\tilde{\phi}^{A}_{I}}{\tilde{\phi}^{\overline{A}}_{J}} & = 0 \\
  \braket{\tilde{\phi}^{A}_{I}}{\tilde{\phi}^{A}_{J}} & = \delta_{IJ} \\
  \braket{\tilde{\phi}^{\overline{A}}_{I}}{\tilde{\phi}^{\overline{A}}_{J}} & = \delta_{IJ} \ .
\end{align}

As $\widehat{n}_{A} + \widehat{n}_{\overline{A}} = \widehat{n}$,
\begin{equation}
  \ket{\widetilde{\phi}_I} = \widehat{n} \ket{\widetilde{\phi}_I} = \widehat{n}_{A} \ket{\widetilde{\phi}_I} + \widehat{n}_{\overline{A}} \ket{\widetilde{\phi}_I} \ ,
\end{equation}
and we can write the single Slater determinant $\ket{\Phi}$ as 
\begin{align} \label{eq:sd_no_basis}
  \ket{\Phi} & = \left( \prod_{I=1}^{N} \widehat{\tilde{a}}_{I} \right) \ket{0} \nonumber \\
  & = \left( \prod_{i=1}^{N} \sqrt{d^A_I} \hat{\tilde{a}}^{A}_{I} + \sqrt{1 - d^A_I} \hat{\tilde{a}}^{\overline{A}}_{I} \right) \ket{0} \ .
\end{align}
Given the above form of $\ket{\Phi}$, the associated density matrix is given by
\begin{equation}
  \widehat{\rho} = \ket{\Phi}\bra{\Phi} = \prod_{I=N}^{1} \hat{\tilde{a}}^{\dagger}_I \prod_{J=1}^{N} \hat{\tilde{a}}_J = \prod_{I=1}^{N} \hat{\tilde{a}}^{\dagger}_I \hat{\tilde{a}}_I \ .
\end{equation}
As the domain-condensed orbital basis of dimension $2N$ is orthonormal, we can construct a Fock space consisting of the the basis $\left\{ \ket{1_{A_I} 0_{\overline{A}_I}}, \ket{0_{A_I} 1_{\overline{A}_I}} \right\}$. In this basis, the density matrix associated with the domain $A$ can be obtained by tracing over the orbitals associated with $\overline{A}$
\begin{equation}
  \widehat{\rho}_{\Omega} = \Tr_{\overline{\Omega}}(\widehat{\rho}) = \Tr_{\overline{\Omega}} (\ket{\Phi} \bra{\Phi}) \rightarrow \bigotimes_{I=1}^{N} 
  \begin{bmatrix}
    1 - d^A_I & 0 \\
    0 & d^A_I
  \end{bmatrix} \ ,
  \label{eqn:density-matrix}
\end{equation}
leading to a density matrix that is expressed solely in terms of the $N$ domain-condensed orbitals of $A$, $\left\{ \ket{1_{A_I}}, \ket{0_{A_I}} \right\}$. This above form of the this density matrix expresses that orbital $I$ is occupied with probability $d^A_I$ and empty with probability $1 - d^A_I$.

\subsubsection{Entanglement measures for atoms: the case of single Slater determinants}

Given the form of the domain density matrix $\rho_{A}$ in \cref{eqn:density-matrix}, the trace of this domain density matrix is given by \cite{song2012a}
\begin{equation}
  \Tr \left( \widehat{\rho}^{\alpha}_{A} \right) = \prod_{I=1}^{N} \Tr \left( \begin{bmatrix}
    \left(1 - d^A_I \right)^{\alpha} & 0 \\
    0 & \left( d^A_I \right)^{\alpha} 
  \end{bmatrix} \right) = 
  \prod\limits_{I=1}^{N} \left( \left(d^A_I\right)^{\alpha} + (1 - d^A_I)^{\alpha} \right) \ ,
\end{equation}
we obtain the following Renyi entropy $S^{A}_{\alpha}$ and von Neumann entropy $S^{A}_{1}$ for the domain $A$ \cite{peschel2003a, calabrese2011a,calabrese2015a} 
\begin{align}
  S^{A}_{\alpha} & = \frac{1}{1-\alpha} \sum\limits_{I=1}^{N} \log( (d^A_I)^{\alpha} + (1-d^A_I)^{\alpha}) \\
  S^{A}_1 & = - \sum\limits_{I=1}^{N} \left( d^A_I \log(d^A_I) + (1 - d^A_I) \log(1 - d^A_I) \right) \ .
\end{align}
Given this entanglement entropy, we can define `information' as the change in entanglement entropy of $A$ as it transitions from being an isolated, closed quantum system to being entangled with its environment \cite{breuer2002a, vanhende2024a}.

If we are interested in correlations and entanglement between $A$ and $B$ in an environment of other atoms then a measure of the total correlations is the quantum mutual information
\begin{equation}
  I_{A:B} = S^{A}_1 + S^{B}_1 - S^{AB}_1 \ ,
\end{equation}
where $S^{AB}_1$ can be constructed by diagonalizing the joint overlap matrix $\Sigma^{\text{occ},AB} = \Sigma^{\text{occ},A} + \Sigma^{\text{occ},B}$. This mutual information quantifies the difference in uncertainty on the quantum state of atom $A$ and the uncertainty on the quantum state of the atom $A$ given that we know the quantum state of the atom $B$. We have recently shown that quantum information analyses in terms of the entanglement entropy and the mutual information capture key properties of quantum atoms and how they interact with their molecular environment \cite{vanhende2024a}.

\subsubsection{Operational considerations lead to superselection rules}

From the viewpoint of quantum information applications, only `accessible' entanglement can be used as a resource \cite{wiseman2003a, klich2008a}. This experimentally accessible entanglement is limited by constraints on local operations (which are also called super-selection rules \cite{ding2020a}) and therefore the amount of accessible entanglement is less than the entanglement entropy. As a local measurement cannot change the particle number of a subsystem, particle number super-selection rules (N-SSR) can be imposed by projecting the density matrices onto fixed particle number sectors in the atoms $A$ and its complement $\overline{A}$ \cite{klich2008a}
\begin{equation}
  \rho_{n,m} = \frac{1}{p_{n,m}} \Pi^{A}_n \otimes \Pi^{\overline{A}}_m \rho \Pi^{A}_n \otimes \Pi^{\overline{A}}_m \ ,
\end{equation}
with 
\begin{equation}
  p_{n,m} = \Tr \left( \Pi^{A}_n \otimes \Pi^{\overline{A}}_m \rho \Pi^{A}_n \Pi^{\overline{A}}_m  \right) \ .
\end{equation}
The accessible or resource entropy of atom $A$ can then be defined as \cite{klich2008a}
\begin{equation}
  S^{\text{acc}}_{A} = \sum\limits_{n=0}^{N} p(n, N-n) \Tr(\rho^{A}_n \log \rho^{A}_n) \ ,
\end{equation}
where 
\begin{equation}
  \rho^{A}_n = \frac{1}{p(n, N-n)} \Pi^{A}_n \rho^{A} \Pi^{A}_n \ .
\end{equation}
As such 
\begin{align}
  S^{\text{acc}}_{A} & = - \sum\limits_{n=0}^{N} \Tr( \Pi^{A}_n \rho^{A} \Pi^{A}_n) \log \frac{\Pi^{A}_n \rho^{A} \Pi^{A}_n}{p(n, N-n)} \\
  & = - \Tr \rho^{A} \log \rho^{A} + \sum\limits_{n=0}^{N} p(n,N-n) \log p(n,N-n) \\
  & = S^{A}_1 - H^{A}_1 \ ,
\end{align}
where
\begin{equation}
  H^{A}_1 = - \sum\limits_{n=0}^{N} p(n, N-n) \log p(n, N-n) \ .
\end{equation}
where $H^{A}_1$ is the Shannon entropy of what is called the electron number distribution function in quantum chemistry \cite{cances2004a, pendas2007a, francisco2014a, pendas2019a} and full counting statistics in quantum physics \cite{nazarov2012a, song2012a,eisler2013a}. This electron number distribution function contains the probability of finding $n$ (and only $n$) electrons in atom $A$ while $N-n$ electrons are found in the environment $\overline{A}$. The associated Shannon entropy was the object of interest in the theory of loges by Daudel \cite{aslangul1972a, aslangul1974a}. In the theory of Maximum Probability Domains by Savin \cite{savin2001a,scemama2007a} one of the probabilities (e.g. $p(2,N-2)$) is maximized by deforming the domain. The resulting MPDs have been shown to provide a quantum chemical correspondence to the Lewis theory of electron pairs \cite{lewis1933a, acke2016a} and to provide support for Clar's aromatic-sextet rule in correlated regimes \cite{vanhende2022a}. Furthermore, the Fourier transform of this probability distribution leads to generating functions which detail a hierarchy of chemical descriptors \cite{acke2020a}.

\section{Thought experiments}

In this section we illustrate the impact of the above derivations using thought experiments that have been explicitly chosen to demonstrate the inconsistency of information theoretical analyses based on the electron density.

\subsection{The case of fully occupied (spin-)position orbitals}

Suppose that a certain position orbital $\ket{\vb{r}}$ is \emph{always} occupied by two electrons, i.e. $p_{11,\vb{r}} = 1$. Then, the electron density at the position orbital $\ket{\vb{r}}$, $\rho^{(1)}(\vb{r})$, is equal to two, the average occupation of the position orbital $\ket{\vb{r}}$, and has \emph{no} probabilistic interpretation. The associated shape function at the position orbital $\ket{\vb{r}}$ is equal to $\sigma^{(1)}(\vb{r}) = \frac{2}{N}$. As such, when there are \emph{only} two electrons, $\sigma^{(1)}$ at the position orbital $\ket{\vb{r}}$ indeed has a the same value as $p_{11,\vb{r}}$. However, if there are more than two electrons in the total system, the shape function decreases accordingly and loses any probabilistic interpretation related to occupation of the position orbital $\ket{\vb{r}}$.

As in this case the one-position-orbital density matrix $\vb*{\rho}$ at $\ket{\vb{r}}$ is given by
\begin{equation}
  \vb*{\rho}_{\ket{\vb{r}}} = \begin{bmatrix}
    0 & 0 & 0 & 0 \\
    0 & 0 & 0 & 0 \\
    0 & 0 & 0 & 0 \\
    0 & 0 & 0 & 1 
  \end{bmatrix} \ ,
\end{equation}
the associated von Neumann entropy $S(\hat{\rho}_{\ket{\vb{r}}})$ is
\begin{equation}
  S(\hat{\rho}_{\ket{\vb{r}}}) = - \sum_{i \in \left\{00, 10, 01, 11 \right\}} p_{i,\vb{r}} \ln(p_{i,\vb{r}}) = 0 \ .
\end{equation}
As such, there is no entanglement between the position orbital and the remainder of the (molecular) system. In contrast, 
\begin{equation}
  - \rho^{(1)}(\vb{r}) \ln \rho^{(1)}(\vb{r}) = - 2 \ln 2 \ ,
\end{equation}
is a (negative) quantification of the \emph{amount of electronic charge} at position-orbital $\ket{\vb{r}}$. The shape function 
\begin{equation}
  - \sigma^{(1)}(\vb{r}) \ln \sigma^{(1)}(\vb{r}) = - \frac{2}{N} \ln \frac{2}{N} \ ,
\end{equation}
is zero for $N=2$, rises to a maximum with increasing number of electrons and then drops uniformly to a limit of zero. This behavior is in spite of, as shown above, there being no entanglement between the position orbital and the remainder of the system.

Suppose that a certain spin-position orbital $\ket{\vb{r}\sigma}$ is always occupied by one electron, i.e. $p_{1,\vb{r}\sigma} = 1$. The spin-resolved electron density $\rho^{(1)}(\ket{\vb{r}\sigma})$ is also equal to one, but integrates to the total number of electrons $N$. If that number of electrons increases $N>1$, then the shape function loses the probabilistic interpretation associated with $p_{1,\vb{r}\sigma}$ as $\sigma^{(1)}(\vb{r}\sigma) = \frac{\rho^{(1)}(\vb{r}\sigma)}{N}$. Although $ -\rho^{(1)}(\vb{r}\sigma) \ln \rho^{(1)}(\vb{r}\sigma) = 0$, a similar construct for the shape function again shows rising and then falling behavior with an increasing number of electrons.

\subsection{A hydrogen molecule at dissociation}

As a thought experiment inspired by Ding et al. \cite{ding2020a}, let us consider the ground state of the hydrogen molecule in the dissociation limit 
\begin{equation}
  \ket{\Psi} = \frac{1}{\sqrt{2}} \left( s^{\dagger}_{L\uparrow} s^{\dagger}_{R\downarrow} + s^{\dagger}_{L\downarrow} s^{\dagger}_{R\uparrow} \right) \ket{0}
\end{equation}
The total correlation between orbitals $\ket{s_L}$ and orbitals $\ket{s_R}$ is equal to $2 \ln 2$. The correlation in the atomic populations is equal to $(1 \ln 1) + (1 \ln 1) = 0$, which implies that if we were to interpret the formulas of Nalewajski and Parr \cite{nalewajski2000a} as information theoretic quantities, the atoms in a statically correlated/maximally entangled hydrogen molecule would have to be considered uncorrelated.

\section{Clearing up misconceptions}

In this section we will discuss well-chosen statements from the literature that are either characteristic of the deep-rooted confusion between densities and probability distributions or are commonly misinterpreted. 

\begin{itemize}

\item \say{$\rho^{(1)}(\vb{x}) \dd \vb{x}$ is the probability of finding a particle with variables $\vb{x}$ in the range $\dd \vb{x}$ at point $\vb{x}$ in configuration space. It should be noted that $\rho^{(1)}(\vb{x})$ integrates to $N$ (not $1$), and that it is therefore the \say{number density}.}\cite{mcweeny1960a}

Although $\rho^{(1)}(\vb{x}) \dd \vb{x}$ \emph{can} be the probability that the orbital $\vb{x}$ is occupied, this is only the case if $\vb{x}$ fully specifies a point in configuration space (i.e. including spin). In that case, there are only two occupations (occupied and unoccupied) and the number density is equal to the probability mentioned above. Although $\rho^{(1)}(\vb{r})$ is still a number density, it is no longer a probability.

\item \say{The electron density quantifies the probability of observing an electron at a point in space, so it is conceptually appealing to define the probability of observing an electron on an atom (i.e., the atomic population) using only the electron density.}\cite{heidar-zadeh2018a}

The electron density is the number density and gives the average occupation of a position orbital. The atomic population is a (weighted) integral over this average occupations and is a population/charge. The probability of `observing an electron at a point in space' or `on an atom' is ill-defined. Only by taking holes explicitly into account can we obtain probability distributions, which in the case of a point in space amounts to the diagonal of the one-position-orbital reduced density matrix and in the case of the atom amounts to the electron number distribution function.

\item 
According to Liu and coworkers \cite{rong2020a}, the first approximation to the electron density information gain 
\begin{equation}
  I_G = \sum\limits_{A} \int \rho_A \ln \frac{\rho_A}{\rho^0_A} \dd \vb{r} \ ,
\end{equation}
is given by 
\begin{equation}
  I_G \approx \sum\limits_A \int (\rho_A(\vb{r}) - \rho^{0}_A(\vb{r})) \dd \bm{r} = -\sum\limits_A q_A = 0
\end{equation}
where $q_A$ is the Hirshfeld charge on A. According to Liu et al., \say{This vanished information gain result suggests that under the first-order approximation, the information before and after a system is formed should be conserved. This first-order approximation result is called in the literature the information conservation principle.}

By misinterpreting the electron density as a probability distribution, Liu et al. mistake the conservation of \emph{charge} with an information conservation principle. A quantum informational analysis also points to a conservation principle, but only when the environment of the proatoms is taken into account. Indeed, at first sight, there seems to be a paradox between considering two hydrogen atoms at infinite separation as an isolated dissociated hydrogen molecule
\begin{equation}
  \frac{1}{\sqrt{2}} \left( \sigma^{\dagger}_{L\uparrow} \sigma^{\dagger}_{R\downarrow} + \sigma^{\dagger}_{L\downarrow} \sigma^{\dagger}_{R\uparrow} \right) \ ,
\end{equation}
with a maximal entropy of $2 \ln 2$, or as two proatoms 
\begin{equation}
  \frac{1}{\sqrt{4}} \left( \sigma^{\dagger}_{L\uparrow} + \sigma^{\dagger}_{R\downarrow} \right) \left( \sigma^{\dagger}_{R\uparrow} + \sigma^{\dagger}_{R\downarrow} \right) \ ,
\end{equation}
with zero von Neumann entropy. The resolution of this paradox is given by Ding et al. \cite{ding2020b}: during dissociation any infinitesimal perturbation will transfer the information contained in this entanglement to the environment of the hydrogen molecule, turning coherent superpositions of degenerate configuration states into classical mixtures of them (i.e. proatoms). As such, in the proatomic viewpoint information can be shown to be conserved, but only when taking the environment explicitly into account (see thermodynamic arguments Ding et al. \cite{ding2020b}). The underlying intuition is that information \emph{has} to be conserved as quantum mechanics is unitary. As the evolution of a subsystem in an environment is not (necessarily) unitary, unitarity can only by restored by taking the environment into account.
\end{itemize}

\section{Conclusions}

In this study, we have shown using quantum information theoretical constructs that the electron density is \emph{not} a probability distribution function and that it should \emph{not} be interpreted as one. Although electron density similarity measures \emph{can} be devised based on information theoretical constructs such as the asymmetric Kullback-Leibler divergence, the resulting numbers do \emph{not} have an information theoretic interpretation. They do not quantify resources that can be used for quantum information processing tasks, but rather quantify the similarity between charge distributions. As the electron density does not obey the axioms that underlie (quantum) information theory, the indiscriminate use of electron densities in a information theoretical framework should be abandoned. Furthermore, previous statements regarding the information content of molecules that are based on the electron density must be reconsidered.

To improve the synergy between conceptual quantum chemistry and quantum information theory, we have discussed in detail the theoretical foundations of electronic (quantum) information theories. We have also shown in detail how a quantum information theory of atoms in molecules can be constructed. The resulting framework clearly indicates that one must go beyond the electron density to construct (quantum) information theories, be it by using a wavefunction ansatz or by constructing practical functionals for higher order reduced density matrices. We explicitly note that this framework is also applicable to ansatzes that are currently used in the quantum simulation of chemistry.

\appendix

\section{Appendix A: Orthogonality of occupied projected domain naturals}\label{sec:orthogonality-orbitals-atoms}

The orthogonality of projected domain naturals can be shown by first expanding two consecutive domain occupation-number operators 
\begin{align}
  & \hat{n}_A \hat{n}_A = \int w_A(\vb{r}) \sum\limits_{\sigma=\uparrow,\downarrow} \hat{\psi}^{\dagger}(\vb{r}\sigma) \hat{\psi}(\vb{r}\sigma) \dd \vb{r} \nonumber \\
  & \quad \quad \quad \quad + \int w_A(\vb{r'}) \sum\limits_{\sigma'=\uparrow,\downarrow} \hat{\psi}^{\dagger}(\vb{r'}\sigma') \hat{\psi}(\vb{r'}\sigma') \dd \vb{r'} \\
  & = \sum\limits_{\sigma=\uparrow,\downarrow} \sum\limits_{\sigma'=\uparrow,\downarrow} \int \int w_A(\vb{r}) w_A(\vb{r'}) \hat{\psi}^{\dagger}(\vb{r}\sigma) \hat{\psi}(\vb{r}\sigma) \hat{\psi}^{\dagger}(\vb{r'}\sigma') \hat{\psi}(\vb{r'}\sigma') \dd \vb{r} \dd \vb{r'} \\
  & = \sum\limits_{\sigma=\uparrow,\downarrow} \sum\limits_{\sigma'=\uparrow,\downarrow} \int \int w_A(\vb{r}) w_A(\vb{r'}) \delta(\vb{r}-\vb{r'}) \delta_{\sigma,\sigma'} \hat{\psi}^{\dagger}(\vb{r}\sigma) \hat{\psi}(\vb{r'}\sigma') \dd \vb{r} \dd \vb{r'} \nonumber \\
  & \quad - \sum\limits_{\sigma=\uparrow,\downarrow} \sum\limits_{\sigma'=\uparrow,\downarrow} \int \int w_A(\vb{r}) w_A(\vb{r'}) \hat{\psi}^{\dagger}(\vb{r'}\sigma') \hat{\psi}^{\dagger}(\vb{r}\sigma) \hat{\psi}(\vb{r}\sigma) \hat{\psi}(\vb{r'}\sigma') \dd \vb{r} \dd \vb{r'} \\
  & = \hat{n}_A - \sum\limits_{\sigma=\uparrow,\downarrow} \sum\limits_{\sigma'=\uparrow,\downarrow} \int \int w_A(\vb{r}) w_A(\vb{r'}) \hat{\psi}^{\dagger}(\vb{r'}\sigma') \hat{\psi}^{\dagger}(\vb{r}\sigma) \hat{\psi}(\vb{r}\sigma) \hat{\psi}(\vb{r'}\sigma') \dd \vb{r} \dd \vb{r'} \ ,
\end{align}
and then applying the resulting operator form to the overlap integral
\begin{align}
  \braket{\tilde{\phi}^{A}_{I}}{\tilde{\phi}^{A}_{J}} & = \frac{1}{\sqrt{d^A_I d^A_J}} \mel{\tilde{\phi}_{I}}{\widehat{n}_{A} \widehat{n}_{A}}{\tilde{\phi}_{J}} \\
  & = \frac{1}{\sqrt{d^A_I d^A_J}} \mel{\tilde{\phi}_{I}}{\widehat{n}_{A}}{\tilde{\phi}_{J}} = \frac{1}{\sqrt{d^A_I d^A_J}} \left( V^{A, \dagger} \Sigma^{\text{occ},A} V^A \right)_{IJ} = \delta_{IJ} \ ,
\end{align}
where the term quartic in field operators drop out as $\ket{\tilde{\phi}_{J}}$ contains only one occupied spin-orbital.

\begin{acknowledgement}

    Parts of this research were funded by the Special Research Fund of Ghent University (Research Project BOF/24J/2019/061). R.V.d.S. acknowledges support from an FWO Ph.D. fellowship (Grant No. 11PN124N).
    
\end{acknowledgement}

\bibliography{first-submission}

\end{document}